\def\ltsima{$\; \buildrel < \over \sim \;$}
\def\lsim{\lower.5ex\hbox{\ltsima}}

\documentclass[
    ,final            
  ]
  {aipproc}

\layoutstyle{6x9}

\begin{document}

\title{What is the radiative process of the prompt phase of Gamma Ray Bursts?}

\classification{98.70.Rz}
\keywords {Gamma Ray Bursts -- radiation processes -- $\gamma$--rays: theory}

\author{G. Ghisellini}{
  address={INAF -- Osservatorio Astronomico di Brera, Via Bianchi 46 
I--23807 Merate, Italy}
}

\begin{abstract}
Despite the dramatic improvement of our knowledge of the 
phenomenology of Gamma Ray Bursts, we still do not know
several fundamental aspects of their physics.
One of the puzzles concerns the nature of the radiative process 
originating the prompt phase radiation. 
Although the synchrotron process qualifies itself as a natural
candidate, it faces severe problems, and many efforts
have been done looking for alternatives.
These, however, suffer from other problems, and there is no
general consensus yet on a specific radiation mechanism. 
\end{abstract}

\maketitle


\section{Introduction}

The field of Gamma Ray Bursts (GRBs) was surrounded by many
years by a sort of fascinating ``aura", being for so long
a complete enigma.
Recent years saw a dramatic improvement of our knowledge
about them, especially about their phenomenology, thanks
to the data gathered by the {\it Compton Gamma Ray Observatory (CGRO)} 
\cite{fishman1995}, {\it Beppo}SAX \cite{frontera2000},
{\it HETE II} \cite{lamb2004}, {\it Swift} \cite{gehrels2009}
and now {\it Fermi} \cite{atwood2009} satellites.

The theoretical work, especially in the 90', 
set the stage for what is now considered a basic standard model
to explain the bulk of what we see (for reviews see e.g.
\cite{vanparadijs2000},
\cite{meszaros2002},
\cite{zhang2004}, \cite{piran2004}, \cite{meszaros2006}).

According to this standard scenario, a colossal injection of energy in a small 
volume lasts for a short time. 
The gravitational energy of a solar mass is liberated in a few seconds, 
in a volume having a radius of a few Schwarzschild radii.
Black--body temperatures above $10^{10}$ K are then reached, and electron--positron
pairs are produced.
The mixture of photons and matter -- the fireball -- expands due to its
internal pressure accelerating the fireball to relativistic velocities.
The Lorentz factor increases as
$\Gamma\propto R$ ($R$ is the distance from the black hole)
until almost all the internal energy is converted into bulk kinetic motion.
A little ``fossil" radiation remains, but it carries a small fraction 
of the initial energy.
There is then the need to reconvert the kinetic energy back to 
radiation. 
The fact that the spikes of emission during the prompt phase do not 
lengthen with time suggests that these episodes occur at the 
same distance from the black hole.
Inhomogeneities in the jet, with regions going at different
$\Gamma$--factors, produce shocks internal to the relativistic flow.
These shocks accelerates electrons and enhance magnetic fields.

Synchrotron radiation is then the natural candidate to explain
the radiation of the prompt phase.
But it faces a severe problem: if electrons produce $\sim$MeV synchrotron photons
with a reasonable efficiency, they must inevitably cool in a 
time \cite{ghisellini2000}:
\begin{equation}
t_{\rm cool} \, =\, 10^{-7} \, { \epsilon_{\rm e}^3 (\Gamma^\prime -1)^3
(\Gamma/100) \over
\nu_{\rm MeV}^2 (1+U_{\rm r}+U_{\rm B}) (1+z)}\,\,\,\, {\rm s}
\end{equation}
where $U_{\rm r}$ and $U_{\rm B}$ are the radiation and magnetic energy
densities, $\epsilon_{\rm e}$ is the fraction of the dissipated energy
given to electrons, and $\Gamma^\prime$ is the relative Lorentz factor
between two colliding shells. 
This time is shorter than any conceivable dynamical time and of any 
detector exposure time.

\begin{figure}
\begin{tabular}{cc}
\includegraphics[height=8.cm, width=7.5cm]{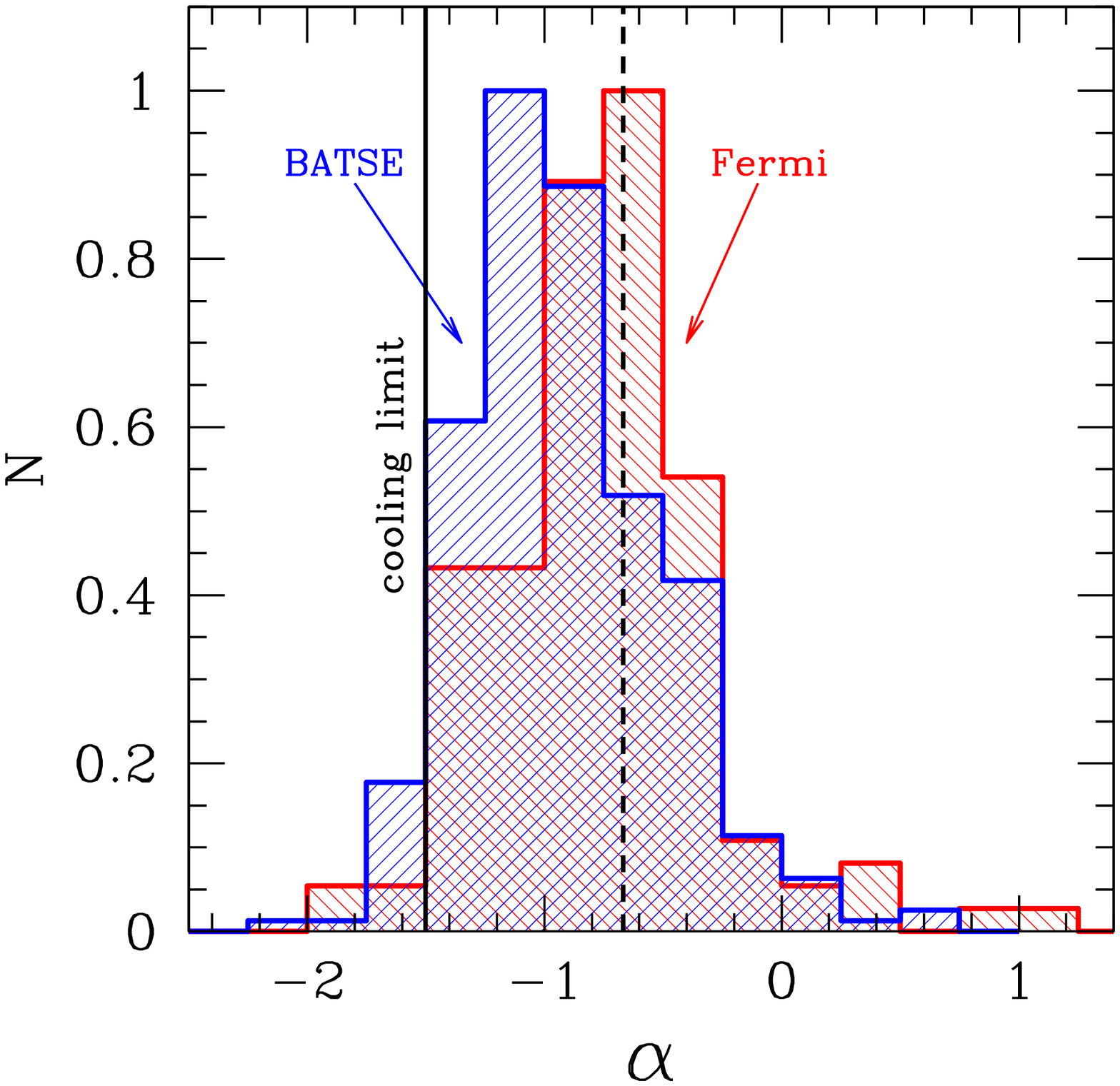}
& \includegraphics[height=8.cm, width=7.5cm]{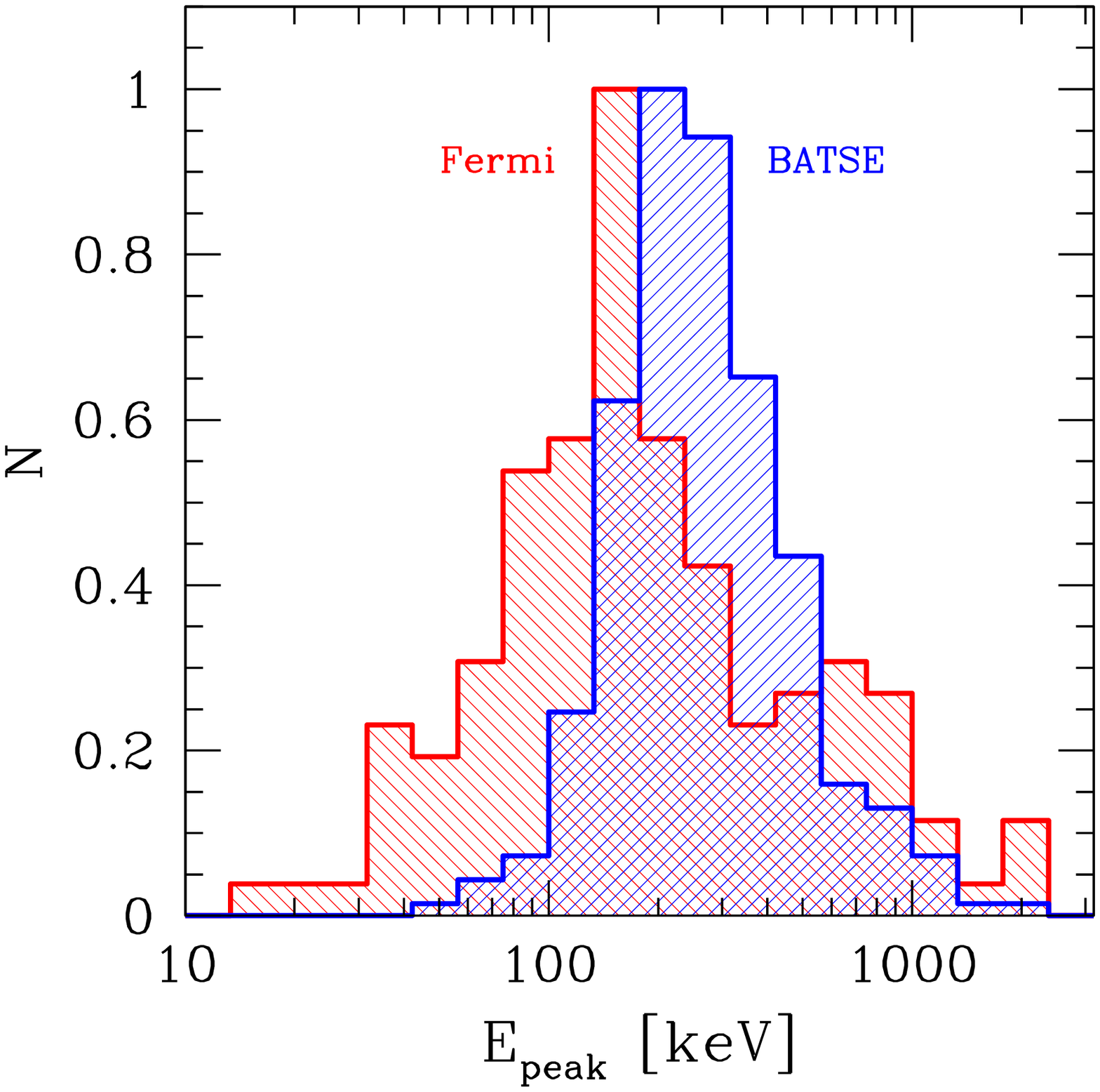}\\
\end{tabular}
\caption{The distributions of low energy spectral indices
$\alpha$ (left) and peak energy $E_{\rm peak}$ (right) 
of BATSE and {\it Fermi}/GBM bursts. 
The spectral index $\alpha$ is the photon spectral index of the
spectrum below the peak energy $E_{\rm peak}$.
The vertical line (left panel) shows the cooling limit ($\alpha=-3/2$), while the
dashed line shows the low energy synchrotron slope of a non cooling electron
population with a low energy cut--off ($\alpha=-2/3$).
Adapted from \cite{nava2010}.
}
\label{isto}
\end{figure}

The synchrotron spectrum of a cooling population of relativistic electrons 
cannot be harder than $F(\nu)\propto \nu^{-1/2}$, corresponding
to a photon spectrum $\dot N(\nu)\propto \nu^{-3/2}$, while the vast majority
of the observed spectra, below their peaks, is much harder, as illustrated
by Fig. \ref{isto}, showing BATSE (onboard {\it CGRO}) and the recent
{\it Fermi} results \cite{nava2010}.
This slope is substantially softer than the 
``synchrotron line of death" \cite{preece1998},
$\dot N(\nu)\propto \nu^{-2/3}$, of a non--cooling electron population
with a low energy cut--off.

\section{Seeking alternatives}

{\bf Re--acceleration --}
The first obvious possibility coming in mind is that the electrons 
are re--accelerated, so that they can remain at the same energy.
In the internal shock scenario this is not possible, since 
electrons are kicked to high energies only once.
Another critical problem is about the global energy budget.
If I keep the radiating electrons hot by refilling their energy,
I can do so with a few (not all) electrons present.
In standard conditions, I can do so only for one in a million electron.

\vskip 0.2 cm
\noindent
{\bf Jitter radiation --}
Small scale changes in direction of the magnetic field can induce the so--called
jitter radiation, similar but not identical to the synchrotron one.
However, if the process is efficient, as it should be, the electrons
cool, and the predicted spectrum is steep.

\vskip 0.2 cm
\noindent
{\bf Self Compton --}
The cooling is very fast anyway \cite{ghisellini2000}, and the predicted first order 
self Compton spectrum is even steeper than the synchrotron one:
$F_{\rm SC}(\nu)\propto \nu^{-3/4}$ in the Thomson regime.
If most of the scatterings occur in the 
Klein Nishina regime, then the electron distribution may flatten,
and the synchrotron radiation is harder \cite{bosnjak2009}, 
but this necessarily implies that the self Compton process is dominating,
in the GeV energy band and beyond.
The few (\lsim 10\%) GRBs detected by {\it Fermi}/LAT at high energies
then suggests that this process is not a general solution.

\vskip 0.2 cm
\noindent
{\bf External Compton --}
The seeds for the Compton process can be 
``fossil" photons remaining from the acceleration phase, or any other
radiation produced externally to the jet.
But the electrons will cool also in this case, if the process is efficient,
making a $\nu^{-1/2}$ spectrum.

\vskip 0.2 cm
\noindent
{\bf Quickly decaying magnetic fields --}
Electrons quickly going away from the acceleration site could emit
in a region of smaller magnetic field, then reducing their
synchrotron losses (and power).
The observed synchrotron spectrum is produced when the
electrons are ``young" and not cooled.
However, once they are out of the magnetised region, 
they would inevitably and efficiently cool by self Compton, that is bound
to become the dominant process, with a corresponding
steep spectrum \cite{ghisellini2000}.

\vskip 0.2 cm
\noindent
{\bf Adiabatic losses ---}
The cooling time is much shorter than any conceivable dynamical time
of the entire fireball.
On the other hand, we could have many small regions 
expanding quickly enough to make electrons loose energy
by adiabatic, not radiative, losses. 
This may also be accompanied by a decreased magnetic field. 
Needless to say, this process is by construction very inefficient.

\vskip 0.2 cm
\noindent
{\bf Quasi--thermal Comptonization ---}
Keeping the internal shock idea, but abandoning the requirement
that the shock accelerates electrons only once, we can envisage
a scenario were all electrons present in the emitting region
(i.e. the shell) are maintained hot by some unspecified process
\cite{ghisellini1999}.
The equilibrium between heating and cooling fixes the typical
energy of these electrons.
If the heating rate is simply the available dissipated energy divided by the
interaction time between two shells, one arrives to typical electron
energies that are sub--relativistic.
The main radiative process in this case is quasi--thermal Comptonization,
using as seed photons either the self absorbed synchrotron photons
produced by the electrons themselves, or the ``fossil" photons.
The Comptonization parameter $y$ becomes of the order of 10 or so,
large enough to produce a hard spectrum.
Expansion of the fireball during the Comptonization process
may quench the process itself (expansion introduces a general radial
motion for both photons and electrons; but see \cite{lazzati2009} for 
a non expanding case resulting from recollimation).
Furthermore, the typical observed energy peak of the spectrum
could be too high, if the electron ``temperatures" in the comoving frame 
are above $\sim 10$ keV or so.

\vskip 0.3 cm
The above ideas have been proposed to occur within the
internal shock scenario.
In the following I will list more radical ideas, that no longer
assume that the dissipation process is due to 
internal shocks.

\vskip 0.3 cm
\noindent
{\bf Bulk Compton ---} 
The association of long GRBs with supernovae led us 
(\cite{lazzati2000}, \cite{ghisellini2000b}) 
to propose an alternative scenario for the production of the prompt phase 
emission, namely to make use of the dense radiation field
produced by the funnel or the progenitor star (that is about to explode)
or by the young and hot remnants (if the supernova explosion 
precedes the GRB).
The process can be very efficient, especially for large $\Gamma$--factors.
There is no need of shocks and no need of a transfer of energy from 
protons to electrons.
Being so efficient, it is conceivable that the fireball decelerates,
leaving less energy to be dissipated during the afterglow.
This would solve another puzzle concerning GRBs.
One of the problems it faces is that the fireball has a large scattering 
optical depth, and so uses a large fraction (if not all) the seed photons.
Furthermore, variability in this model should correspond to emission by
different shells, but there is a minimum ``refilling time" needed to 
replace the scattered seed photons with new ones.
Also, the similarity of the spectra of long and short GRBs \cite{ghirlanda2009}
(if short bursts are not associated to a supernova) makes
the bulk Compton idea questionable.

\vskip 0.2 cm
\noindent
{\bf Deep impacts ---}
The initial fireball must punch a funnel through the progenitor star.
If the opening angle of the jetted fireball is $\theta =0.1\sim 5^\circ$, then
each one of the two oppositely directed fireballs must push
a mass $M\sim 0.5 (\theta/0.1)^2 (M_*/20\, M_\odot)$ solar masses
out of the progenitor star of mass $M_*$.
Once the funnel is clean, the fireball may still interact with some
material leftover from the previous (``piercing") phase, at a distance
of the same order of the star radius \cite{ghisellini2007}.
Moreover, shear instabilities between the fireball and its cocoon
(while the fireball is moving inside the funnel) may give important dissipation
\cite{thompson2007}, \cite{lazzati2009}.
The efficiency can be large, especially when the fireball
collides with leftover material just outside the star surface,
because it is a collision with matter that is initially almost at rest.
If these collisions occur when the scattering optical depths are large, then
the predicted spectrum has time to thermalize, and then it is a
black--body.
Under the assumption of black--body spectrum and other
specific conditions (even if they appear somewhat ad hoc), 
\cite{thompson2007} showed that it is even possible to reproduce the ``Amati"
relation \cite{amati2002}, namely the correlation between the observed energetics
of the prompt radiation phase and the peak energy of the $\nu F_\nu$ spectrum
of the total prompt emission.

The problem with these interesting attempts is the presence of a black--body
component in the prompt phase spectrum.
While some GRBs do have black--body like spectra up to a few seconds
from the trigger \cite{ghirlanda2003}, the vast majority do not.
Fits with a back--body plus a power law can be successfully 
applied to many more bursts \cite{ryde2005}, but the resulting power law
is rather soft.
Therefore, even if in the BATSE energy range one obtains a good fit,
the extrapolation of the power law component to lower frequencies results
in a large flux. Larger than what observed when we do have lower frequency data,
as was the case for the few GRB observed both by BATSE and by the Wide Field Camera
(WFC) onboard {\it Beppo}SAX \cite{ghirlanda2007}.
Moreover, for all those cases, a cut--off power law (without the black--body component)
not only is a good fit in the BATSE energy range, but also its extrapolation
to lower frequencies matches perfectly the WFC data.

\begin{figure}
\includegraphics[height=7cm, width=16cm]{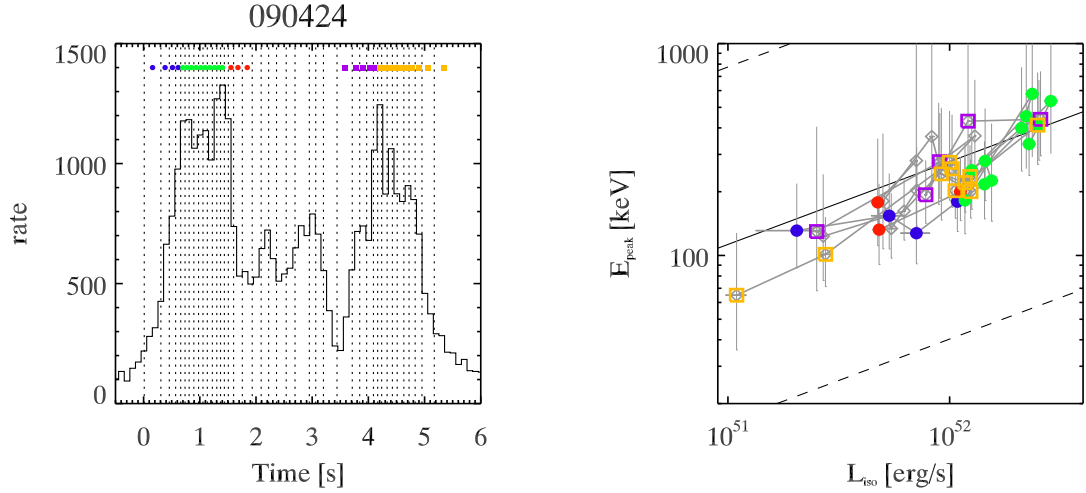}
\caption{The left panel shows the {\it Fermi}/GBM light curve of GRB 090424.
Vertical dashed lines indicate the time bins for 
which the spectrum was analysed.
The right panel shows $E_{\rm peak}$ vs luminosity for the different
time bins. The solid and dashed dark lines indicate the slope and normalisation
of the Yonetoku relation found considering different GRBs.
Different symbols indicate the rising and decaying phases of the different pulses.
Adapted from \cite{ghirlanda2009}.
}
\label{090424}
\end{figure}

\vskip 0.2 cm
\noindent
{\bf Reconnection ---}
The fireball could be magnetically dominated, dissipating 
part of its magnetic energy through reconnection,
as envisaged in \cite{giannios2008}.
If this kind of energy dissipation lasts for a relatively
long time, then the electrons would be
reaccelerated while cooling, and the dominant radiation
process could be similar to the quasi--thermal Comptonization
mechanism.
The several spikes/pulses present in the light curve of the prompt
emission phase should correspond to different reconnection events.
This idea is attractive, and surely worth to be investigated further.
The problem with it is that assuming dissipation events having
different properties (i.e. energies, electron content, sizes, durations)
would correspond to ``Christmas tree" variations, namely each spike/pulse
should behave independently from the others.
There should be no well defined trends in the spectral properties
of the prompt emission.
Instead, these trends are present, and are rather strong.
In fact, both \cite{firmani2009} (for {\it Swift} bursts) and
\cite{ghirlanda2009} (for {\it Fermi}/GBM bursts)
found strong correlations between the peak energy $E_{\rm peak}$ 
and the luminosity {\it within the prompt emission of single bursts}.
Fig. \ref{090424} 
illustrates the point showing the $E_{\rm peak}$--Luminosity
correlation for GRB 090424.
The slope and normalisation of these correlation is the same of
what it is found considering different bursts, and taking for each
of them the peak luminosity and the (time averaged) $E_{\rm peak}$
(the so called Yonetoku correlation, see \cite{yonetoku2004},
and \cite{ghirlanda2009} for an update).

These trends give solidity and reality to
spectral--energy correlations found in these years, demonstrating that 
{\it they are not} 
the result of selection effects.

\section{Conclusion}

We still do not know what is the dominant radiation process
of the prompt phase emission.




\begin{theacknowledgments}
I thank G. Ghirlanda for discussions.
This work was partially funded by a 2007 PRIN--INAF grant.
\end{theacknowledgments}



\bibliographystyle{aipproc}   




\end{document}